\documentclass[%
 reprint,
superscriptaddress,
 amsmath,amssymb,
 aps,
pra,
showkeys
]{revtex4-2}

\usepackage{romannum}
\usepackage{booktabs}
\usepackage{graphicx}
\usepackage{lipsum}
\usepackage{placeins}
\usepackage{amsmath}
\usepackage{hyperref}
\hypersetup{
    colorlinks = true,
    allcolors = {blue},
}

\usepackage{epstopdf}

\newcommand{\andGang}{\textit{et al.}}    

\usepackage{graphicx}
\usepackage{dcolumn}
\usepackage{bm}

\begin{document}

\preprint{APS/123-QED}

\title{Efficient spin-wave excitation by surface acoustic waves in ultra-low damping YIG/ZnO-heterostructures}%

\author{Yannik Kunz} 
\email{ykunz@rptu.de}
\affiliation{Fachbereich Physik and Landesforschungszentrum OPTIMAS, Rheinland-Pfälzische Technische Universität Kaiserslautern-Landau, 67663 Kaiserslautern, Germany}
\author{Julian Schüler}
\affiliation{Fachbereich Physik and Landesforschungszentrum OPTIMAS, Rheinland-Pfälzische Technische Universität Kaiserslautern-Landau, 67663 Kaiserslautern, Germany}
\author{Finlay Ryburn}
\affiliation{Clarendon Laboratory, Department of Physics, University of Oxford, Parks Road, Oxford,
OX1 3PU, United Kingdom}
\author{Kevin Künstle}
\author{Michael Schneider}
\affiliation{Fachbereich Physik and Landesforschungszentrum OPTIMAS, Rheinland-Pfälzische Technische Universität Kaiserslautern-Landau, 67663 Kaiserslautern, Germany}
\author{Katharina Lasinger}
\affiliation{Fachbereich Physik and Landesforschungszentrum OPTIMAS, Rheinland-Pfälzische Technische Universität Kaiserslautern-Landau, 67663 Kaiserslautern, Germany}
\affiliation{Clarendon Laboratory, Department of Physics, University of Oxford, Parks Road, Oxford,
OX1 3PU, United Kingdom}
\author{Yangzhan Zhang}
\affiliation{Clarendon Laboratory, Department of Physics, University of Oxford, Parks Road, Oxford,
OX1 3PU, United Kingdom}
\author{Philipp Pirro}
\affiliation{Fachbereich Physik and Landesforschungszentrum OPTIMAS, Rheinland-Pfälzische Technische Universität Kaiserslautern-Landau, 67663 Kaiserslautern, Germany}
\author{John Gregg}
\affiliation{Clarendon Laboratory, Department of Physics, University of Oxford, Parks Road, Oxford,
OX1 3PU, United Kingdom}
\author{Mathias Weiler}
\affiliation{Fachbereich Physik and Landesforschungszentrum OPTIMAS, Rheinland-Pfälzische Technische Universität Kaiserslautern-Landau, 67663 Kaiserslautern, Germany}

\date{\today}

\begin{abstract}
We demonstrate the efficient excitation of spin waves in the ultra-low magnetic damping material yttrium-iron-garnet (YIG) by surface acoustic waves (SAWs). To this end, we employ interdigital transducers fabricated on a piezoelectric zinc oxide (ZnO) thin film covering the YIG. This enables the excitation of coherent, propagating Rayleigh-type and Sezawa-type SAWs. We find that the ultralow magnetic damping of YIG is retained after the ZnO-deposition and we carry out a comprehensive investigation of the magnetoelastic interaction due to the different SAW modes as a function of the external magnetic field magnitude and orientation. By measuring the SAW attenuation our experiments reveal a highly anisotropic and non-reciprocal SAW-SW-interaction in agreement with our model caluclation, while demonstrating that low-damping magnons can be efficiently excited by SAWs.  
\end{abstract}

\maketitle


\section{\label{sec:Introduction}Introduction}

In recent years, the control and manipulation of magnetization and its' dynamics by magnetoacoustic effects has attracted significant interest~\cite{yangAcousticControlMagnetism2021}. Research topics on surface acoustic wave (SAW) driven magnetoacoustics range from magnetoelastic switching of magnetic states~\cite{Sampath2016}, moving of skyrmions~\cite{yangAcousticdrivenMagneticSkyrmion2024} and domain walls~\cite{Edrington2018} or the excitation of vortex dynamics~\cite{Iurchuk2024, Koujok2023}, to investigations of magnon-phonon coupling strength~\cite{hwangStronglyCoupledSpin2023}, among others. Leveraging hybrid mechanisms such as magnetoelastic~\cite{weilerElasticallyDrivenFerromagnetic2011}, magnetorotational~\cite{xuNonreciprocalSurfaceAcoustic2020, kussNonreciprocalDzyaloshinskiiMoriya2020}, spin-rotational~\cite{kobayashiSpinCurrentGeneration2017} or out-of-plane phononic angular momentum~\cite{liaoNonreciprocalMagnetoacousticWaves2024} coupling paves the way for exploring novel phenomena, both fundamentally and for applications. Recent studies have extended the investigation of magnetoacoustic interactions to synthetic~\cite{Kuess2023, Seeger2024} and single-phase antiferromagnets~\cite{zhangTerahertzFieldinducedNonlinear2024}. The excitation of magnetization dynamics in the form of spin waves (SWs) by acoustic waves, also referred to as phonon-magnon-interaction, is intensively studied. On one hand, the magnetoelastic interaction of SAWs with SWs in complex multilayered systems has been shown to be an effective tool for manipulation of the SAW transmission, giving rise to giant non-reciprocity~\cite{Kuess2023, Kuess2024, Seeger2024, thevenardSurfaceacousticwavedrivenFerromagneticResonance2014}. On the other hand, the SW excitation by SAWs can exceed the non-linear criticality~\cite{geilenParametricExcitationInstabilities2024} of spin-wave stability by parametric amplification~\cite{Jander2025}.  Thus, the excitation of magnons via magnetoacoustic interactions is a promising approach, to overcome the challenge of launching spin-waves efficiently in magnonic devices, which aim to exploit magnons for information processing~\cite{Mahmoud2021} or computing~\cite{Chumak2014, Wang2018, Kostylev2005}. The intrinsic nonlinearity of SWs renders them ideal candidates for novel computing schemes like neuromorphic~\cite{Chumak2022}, reservoir~\cite{Papp2021} and even quantum computing~\cite{yuanQuantumMagnonicsWhen2022, Andrianov2014, Mohseni2022}. Intrinsically, magnonic applications rely on the coherent excitation of SWs~\cite{Pirro2021}, as is achievable by magnon-phonon interactions~\cite{hiokiCoherentOscillationPhonons2022, kunzCoherentSurfaceAcoustic2024a}.\\    
Commonly employed magnetostrictive metallic magnets, such as Ni~\cite{weilerElasticallyDrivenFerromagnetic2011}, Fe~\cite{Duquesne2019} or Py~\cite{Long1966} typically exhibit comparatively high magnetic damping~\cite{Kobayashi2009, Walowski_2008, PhysRevB.81.174414}. While these materials are well-suited for manipulation of magnetic states~\cite{Sampath2016} or modulation of the SAW transmission~\cite{kussNonreciprocalDzyaloshinskiiMoriya2020}, their high Gilbert damping coefficients results in short magnon lifetimes. This renders them unsuitable for magnonic applications, in which long-living spin waves contribute to the energy efficiency of the devices. The typical material of choice in magnonics is Yttrium-Iron-Garnet (YIG)~\cite{gellerStructureFerrimagnetismYttrium1957, cherepanovSagaYIGSpectra1993b}, as it holds record magnon lifetimes. The weak spin-orbit interaction of YIG yields magnetoelastic coefficients that are one order of magnitude smaller compared to typical ferromagnetic metals, rendering YIG unsuitable for devices exploiting the manipulation of the SAW transmission by magnetostriction. However, as the magnetoelastic effective magnetic field scales inversely with the saturation magnetization, with $M_\text{S, YIG}\approx140\;$kA/m~\cite{cherepanovSagaYIGSpectra1993b}, efficient SW excitation via SAWs can still be achieved as we demonstrate in the following. In previous realizations of magnetoacoustic devices based on YIG grown on piezoelectric materials via sputtering deposition or pulsed laser deposition (PLD)~\cite{wongAcousticallyDrivenFerromagnetic2024}, the magnetic damping was vastly increased due to the lattice mismatch between the piezoelectric and the YIG. Although it is possible to deposite YIG with ultra-low magnetic damping by sputtering~\cite{TORRAO2022166300, 9076316} or PLD~\cite{Onbsli:2014}, comparable to that of YIG grown by liquid phase epitaxy (LPE), the gadolinium gallium garnet (GGG) substrate remains indispensable.\\ 
To study magnetoacoustic interactions in low-damping YIG, we here employ heterostructures consisting of a sputter-deposited piezoelectric zinc oxide ZnO layer fabricated on top of a YIG film grown on a GGG subtrate by LPE. Similar heterostructures have been exploited in bulk acoustic resonators~\cite{Alekseev2020}, and for MHz-frequency SAW devices~\cite{lewisAcousticSurfaceWaveIsolator1972, inabaParametricAmplificationSurface1982, voltmerMAGNETOSTRICTIVEGENERATIONSURFACE1969}.  
In contrast to these studies, we here demonstrate the coherent excitation of GHz-frequency magnons by SAW phonons and associated resonant SAW attenuation in YIG in a systematic study of the magnon-phonon interaction as a function of SAW type, and external magnetic field orientation. We further theoretically model the phonon-magnon interaction and provide experimental evidence that the resonant spin-wave excitation occurs while preserving the ultra-low damping YIG properties.\\

\section{Magnon-excitation by Surface Acoustic Waves in YIG/ZnO}

On our device, we employ sets of interdigital transducers (IDTs), patterned on the piezoelectric ZnO to launch  propagating SAWs. The induced periodic elastic deformation extends through the ZnO-layer into the YIG-layer, where the elastic deformation interacts with the spin-system. The sample fabrication and properties of the resulting gigahertz-range SAWs in a comparable layer stack are reported in detail in~\cite{ryburn2024generationgigahertzfrequencysurface}.
\begin{figure}
\includegraphics{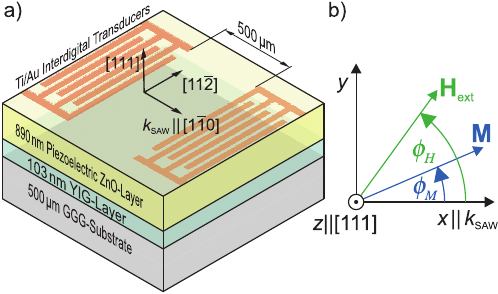}
\caption{\label{fig:Samplesketch}Schematic sketch of the investiagted device and the coordinate system. As shown in Panel a), the sample consists of a GGG(500\;µm)/YIG(103\;nm)/ZnO(890\;nm)-stack with Ti/Au-interditigal transducers of a 2.8\;µm periodicity. The employed coordinate system is illustrated in Panel b), indicating the direction of the in-plane oriented external magnetic field $\textbf{H}_\text{ext}$ and the magnetization $\textbf{M}$ relative to the SAW propagation direction, defined by $\textbf{k}_\text{SAW}$.}
\end{figure}
The investigated device consists a 103\;nm thin YIG layer grown in [111]-direction on a 500\;µm GGG substrate via LPE. An 890\;nm ZnO-layer is deposited via rf-magnetron sputter deposition, while sets of Ti(5\;nm)/Au(70\;nm) IDTs are structured by electron beam lithography and electron beam evaporation. The sending- and the receiving IDTs are located 500\;µm apart, with the SAW propagating parallel to the $[1\bar{1}0]$ direction. A schematic of the sample is shown in Fig.\;\ref{fig:Samplesketch} a) and details on the preparation are provided in the Supplemental Material. The magnetic parameters of the YIG after the ZnO-depositon are determined by out-of-plane- and in-plane-angle-resolved ferromagnetic resonance spectroscopy. The results are summarized in Appendix \ref{App:FMR}. Notably, the extracted Gilbert damping of $\alpha=5.2\times10^{-4}$ is comparable to that found for YIG films of similar thickness without ZnO~\cite{dubsSubmicrometerYttriumIron2017}, indicating the preservation of the ultra-low magnetic damping property of the YIG after ZnO deposition. \\ 
We define our coordinate system such that the x-axis lies in-plane, parallel to the propagation direction of the SAW, while the z-axis is oriented out-of-plane along the [111]-direction. The external field- and magnetization directions are parametrized by their angle relative to the SAW propagation, defined by $\textbf{k}_\text{SAW}$, as illustrated in Fig.\;\ref{fig:Samplesketch} b).\\
To study the SAW-SW interaction we connect the opposing set of IDTs to a vector network analyzer. By measuring the $S_{21}$-transmission parameter and utilizing time-gating~\cite{chenDelvingTimeDomain2024, Wang2018}, we extract the transmitted SAW amplitude. The sample is placed in a rotatable electromagnet, to change the externally applied magnetic field magnitude $\mu_0\textbf{H}_\text{ext}$ and orientation $\phi_H$ relative to the SAW propagation direction.\\
\begin{figure}[t]
\includegraphics{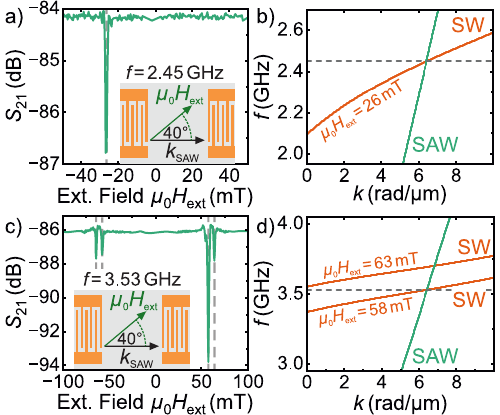}
\caption{\label{fig:Fields_SweepDispRel}Magnetic field dependent SAW absorption for the Rayleigh-type mode at 2.45\;GHz in a) and the Sezawa-type mode at 3.53\;GHz in c) at a fixed external field orientation of $\phi_H=40$°. The corresponding SW and SAW dispersion-relations at the resonance magnetic field is shown in b) and d).}
\end{figure}
\begin{figure*}[t]
\includegraphics{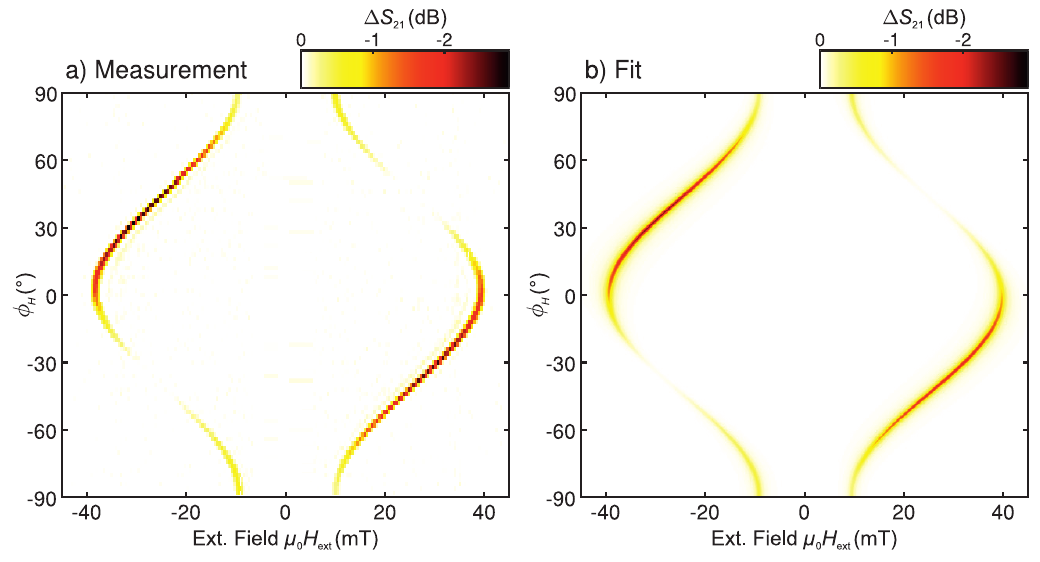}
\caption{\label{fig:MECRayleighMode}Coupling symmetry of the Rayleigh-type SAW mode measured at 2.45\;GHz. a) shows the measurement and b) the theoretical modeling. Remarkably, the interaction is highly non-reciprocal. The used magnetic parameters are determined by a least-squares-optimization using the position of the resonance, while the strain-parameters of the SAW are extracted from an optimization of the full dataset. All parameters used are summarized in Table\;\ref{tab:SAWFitparams}}
\end{figure*}
We first apply the external magnetic field at the fixed orientation of $\phi_H=40$° and investigate the transmission of the Rayleigh-type SAW at a frequency of around 2.45\;GHz and the Sezawa-type SAW at 3.53\;GHz as a function of field strength. The transmission parameter $S_{21}$ is shown in Fig.\;\ref{fig:Fields_SweepDispRel} a) and c). As can be seen, for the Rayleigh-type mode at 2.45\;GHz in Fig.\;\ref{fig:Fields_SweepDispRel} a), we observe a resonant absorption of the SAW at an external field of $\|\mu_0H_\text{ext}\|=26\;\text{mT}$. This field corresponds to the intersection of the SW dispersion with the SAW dispersion shown in Fig.\;\ref{fig:Fields_SweepDispRel} b) at the wave vector of 6.43\;rad/µm, indicating the resonant interaction of the phonons with magnons. Thus, during the interaction the energy and momentum conservation is fulfilled:

\begin{align}
    hf_\text{phonon}=hf_\text{magnon},\\
    \hbar k_\text{phonon}=\hbar k_\text{magnon},
\end{align}
where $k$ denotes the wave vector and $f$ the frequency. Notably, the interaction is highly non-reciprocal at this field orientation, showing a relative absorption of approximately -2.8\;dB at the negative resonance magnetic field of -26\;mT, whilst seemingly no absorption is observed at positive magnetic fields. This is attributed to the helicity-mismatch-effect~\cite{kussNonreciprocalDzyaloshinskiiMoriya2020, yamamotoInteractionSurfaceAcoustic2022}, which yields differing interaction efficiencies, depending on the helicity of the SAW and the the precession orientation of the SWs match. This non-reciprocity will be discussed in the context of our angle-resolved measurements later on.\\
\begin{figure*}[t]
\includegraphics{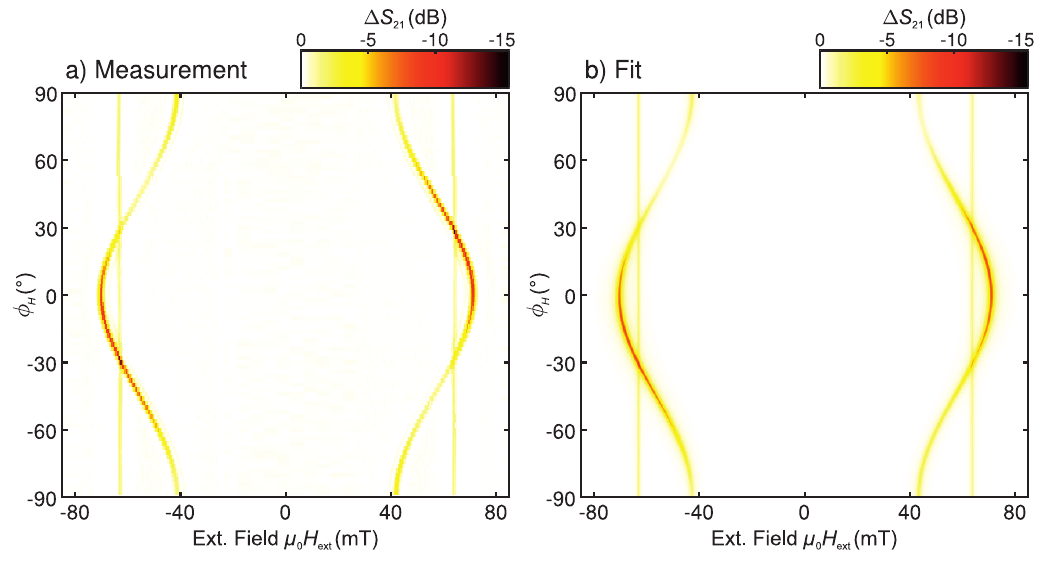}
\caption{\label{fig:MECSezawaMode}Coupling symmetry of the Sezawa-type SAW mode measured at 3.53\;GHz. a) shows the measurement and b) the theoretical modeling. Notably, the interaction symmetry is roughly inverted compared to the Rayleigh-type mode shown in Fig.\;\ref{fig:MECRayleighMode} as a consequence of the differing strain components and reversed helicity. The angle-independent absorption line at $53\;$mT results from a direct excitation of the ferromagnetic resonance.}
\end{figure*}
In the case of the Sezawa mode at $f=3.53\;$GHz, shown in Fig.\;\ref{fig:Fields_SweepDispRel} c), we find multiple absorption lines, at the fields of $\|\mu_0H_\text{ext}\|=58\;\text{mT}$ and at $\|\mu_0H_\text{ext}\|=63\;\text{mT}$. Here, the resonance at higher field is again found at the intersect of the SAW and the SW dispersion relations. Additionally, the lower-field resonance occurs around the field of the $k=0$-mode, i.e. the ferromagnetic resonance (FMR). The FMR excitation is caused by a direction excitation of the spin-system by the ac-current in the microstructures, as we verify and discuss in Appendix \ref{App:S11}. Notably, for the Sezawa-mode the absorption is most prominent at positive values of the externally applied magnetic field, opposite to the corresponding observations for the Rayleigh-type mode in Fig.\;\ref{fig:Fields_SweepDispRel} a). This is attributed to the differing contributions of individual strain-component components of the Rayleigh-type and the Sezawa-type modes, in agreement with their reversed helicity~\cite{Cheeke2002}.

\section{Magnetoacoustic Interaction Symmetry of Rayleigh- and Sezawa type modes}

In order to map the magnetoacoustic interaction symmetry driven of the two SAW modes, we perform field- and angle-dependent measurements. To focus on the magnetic field dependence, we consider the relative SAW transmission:
\begin{equation}
    \Delta S_{21}(\mu_0H_\text{ext})= S_{21}(\mu_0H_\text{ext}) - \bar{S}_{21}(\|\mu_0H_\text{ext}\|>40\;\text{mT})
\end{equation}
where $\bar{S}_{21}$ denotes the averaged transmission parameters at fields $\|\mu_0H_\text{ext}\|>40\;\text{mT}$. The resulting relative SAW transmission is shown in Fig.\;\ref{fig:MECRayleighMode} a). The SAW-SW interaction is highly anisotropic and non-reciprocal. The anisotropy is accounted to the anisotropic dipolar-exchange SW dispersion which dominates the angular dependence of the resonant magnetic field, while the magnetic anisotropy is only in the order of about $\mu_0H_\text{ani}=-2\mu_0K_\text{u}/M_\text{S}\approx0.73\;$mT as shown in Appendix \ref{App:FMR}.  Lewis and Patterson~\cite{lewisAcousticSurfaceWaveIsolator1972} reported a non-reciprocal SAW attenuation at a frequency of 200\;MHz and a field orientation of $\phi_H=90$°. In their study the origin of the arising non-reciprocity could not be clearly assigned and the non-reciprocity occurred under an angle where the SAW-SW interaction is reciprocal in our data. In Fig.\;\ref{fig:MECRayleighMode} a), the Rayleigh-mode has a non-vanishing coupling efficiency at $\phi_H=0$° and 90°, corresponding to the Damon-Eshbach and the Backward-volume spin-wave geometry. This warrants a closer investigation of the symmetry of the magnetoacoustic interaction, as no magnetoacoustic coupling is expected for these orientations under the assumption of a pure compressional SAW mode \cite{kussSymmetryMagnetoelasticInteraction2021}.\\
In order to model the measurement data, we follow the approach developed by Dreher \andGang~\cite{Dreher2012} and Küß \andGang~\cite{kussNonreciprocalDzyaloshinskiiMoriya2020}, employing the effective magnetic driving field approach, detailed in Appendix \ref{App:MEC}. The total absorbed Power $P_\text{abs}$ is derived by:
\begin{equation}
    P_{\text{abs}}=-\mu_0\pi f\text{Im}\left(\int_V \textbf{h}_\text{MEC}^*\chi \textbf{h}_\text{MEC}\text{d}V\right)\;,
\end{equation}
where $\chi$ is the dynamical Polder susceptibility tensor, $f$ the frequency, $\textbf{h}_\text{MEC}$ the magnetoacoustic excitation field and $V$ the magnetic volume. For the derivation of $\textbf{h}_\text{MEC}$ see Appendix \ref{App:MEC}. We determine the magnetic parameters by first fitting the angle-dependent resonance magnetic field positions with the spin-wave dispersion. For this, the thin film approximation by Kalinkos-Slavin for description of the dipolar terms, the exchange interaction terms and the anisotropy contributions are included~\cite{kalinikosTheoryDipoleexchangeSpin1986}. Under the assumption that $\textbf{M}\|\textbf{H}$, the SW-dispersion can be expressed as:
\begin{equation}
    f=\frac{\gamma}{2\pi}\mu_0\sqrt{H_{11}H_{22}}.
\end{equation}
where $f$ denotes the frequency, $\gamma$ is the gyromagnetic ration and $H_{11}$ and $H_{22}$ terms depend on the external field and the contributions mentioned above (see Appendix \ref{App:MEC}). A summary of the used magnetic parameters is given in Table\;\ref{tab:SAWFitparams}.\\
\begin{table}[b]
\caption{\label{tab:SAWFitparams}%
Magnetic parameters and normalized strain components of both the Rayleigh- and the Sezawa-mode obtained by the parameter optimization. The magnetoelastic coupling coefficients $b_1$ and $b_2$ are takten from Ref.~\cite{comstockMagnetoelasticCouplingConstants1965}. For the definition of the strain component amplitude and phase provided below, please see the supplemental material.}
\begin{ruledtabular}
\begin{tabular}{lll}
Quantity & \multicolumn{2}{c}{Mode} \\
& \textrm{Rayleigh} & \textrm{Sezawa}\\
\colrule
Frequency $f$ (GHz) & 2.45 & 3.53 \\
Group velocity $v_\text{G}$ (m/s) & 2458 & 2641 \\
Wave vector $k$ (rad/µm) & \multicolumn{2}{c}{6.43} \\
Film thickness YIG $d$ (nm) & \multicolumn{2}{c}{94.2} \\
g-factor $g$ & \multicolumn{2}{c}{2.11} \\
Sat. magnetization $\mu_0M_\text{S}$ (mT) & \multicolumn{2}{c}{0.158} \\
Exchange constant $A$ (pJ/m) & \multicolumn{2}{c}{3.41} \\
Uniaxial aniso. field $\mu_0H_\text{u}$ (mT) & \multicolumn{2}{c}{-0.365} \\
Uniaxial aniso. easy axis $\phi_\text{u}$ (°) & \multicolumn{2}{c}{-43.3} \\
OOP aniso. field $\mu_0H_\text{OOP}$ (mT) &  \multicolumn{2}{c}{-7.0} \\
Effective damping $\alpha_\text{eff}$ & 0.0070 & 0.0069 \\
Magn.-elas. Coup. Coeff. $b_1\;$(J/m\textsuperscript{-3}) &  \multicolumn{2}{c}{$3.48\times10^{-7}$}\\
Magn.-elas. Coup. Coeff. $b_2\;$(J/m\textsuperscript{-3}) &  \multicolumn{2}{c}{$6.96\times10^{-7}$} \\
Amplitude $\varepsilon_\text{xx}$   & 1             & 0.438 \\
Phase $\varepsilon_\text{xx}$ (rad) & 0             & 0 \\
Amplitude $\varepsilon_\text{xy}$   & 0.186    & 0.306 \\
Phase $\varepsilon_\text{xy}$ (rad) & -1.089\;$\pi/2$   & -1.218\;$\pi/2$\\
Amplitude $\varepsilon_\text{xz}$   & 0.470    & 1 \\
Phase $\varepsilon_\text{xz}$ (rad) & 0.944\;$\pi/2$   & -1.177\;$\pi/2$ \\
Amplitude $\varepsilon_\text{yz}$   & 0.062    &  0.166 \\
Phase $\varepsilon_\text{yz}$ (rad) & 1.324\;$\pi/2$   & 0.586\;$\pi/2$
\end{tabular}
\end{ruledtabular}
\end{table}
We treat the relative strain components that define the SAW type as free parameters, they are determined by a least-squares optimization of the experimental data. Hereby, a set of normalized strain parameters is used, and the total absorption is multiplied by a scaling parameter. Further, the effective Gilbert-damping parameter (definition see Appendix \ref{App:MEC}, which parametrizes the line-width is fitted. The parameters obtained are provided in Table\;\ref{tab:SAWFitparams}. As can be seen in Fig.\;\ref{fig:MECRayleighMode} b), the theoretical model with the set of optimized parameters can quantitatively reproduce the experimental data set.\\
Next, we focus on the investigation of the Sezawa-like mode at $f=3.53\;$GHz and its SAW-SW interaction symmetry. We again measure the relative SAW transmission in dependence of magnetic-field and orientation, now choosing for normalization of $\Delta S_{21}$ the field range $\|\mu_0H_\text{ext}\|>80\;$mT. The measurement results obtained are shown in Fig.\;\ref{fig:MECSezawaMode} a). As expected, at the higher frequency the resonance condition shifts towards larger external magnetic fields. The angle-resolved measurement reveals the full coupling symmetry. As expected, the maximum coupling efficiency is now observed at $\phi_H$-values that differ from those observed for the Rayleigh-mode shown in Fig.\;\ref{fig:MECRayleighMode}. 
The difference in the interaction symmetry arises from the reversed helicity of the SAW, i.e., its polarization, relative to the Rayleigh mode. This reversal occurs due to an additional strain node within the material, as the Sezawa mode represents a higher-order SAW mode compared to the Rayleigh mode. This becomes apparent in Table\;\ref{tab:SAWFitparams} by the sign change of the $\varepsilon_\text{xz}$-phase component. In Fig.\;\ref{fig:MECSezawaMode} a), a near vertical line of contrast is visible at an external magnetic field of approximately $\mu_0H_\text{ext}=58\;$mT. We argue that this additional excitation does not originate from the interaction of the SAW with the spin-system, but is caused by excitation of the uniform SW mode (FMR). The FMR field is independent of $\phi_H$, in accordance with vanishing in-plane magnetic anisotropy. We attribute the FMR excitation to the ac-current in the microstructured pads, as the momentum-conservation between the phonon and the FMR-magnon could not be fulfilled. Further, the direct FMR excitation is continuous in frequency and therefore not constrained by the IDT periodicity, which limits the excitation spectra of the IDTs to discrete frequencies. The measurement of the $S_\text{11}$-reflection parameters confirms this, as shown in Fig.\;\ref{fig:S11Reflec} in Appendix \ref{App:S11}. Further, we observe a large increase of the absorbed SAW power from around 2.8\;dB for the Rayleigh-mode, to around 15\;dB for the Sezawa-mode. On one hand, this can be attributed to the increase in frequency, resulting in an increased power absorption as $P_\text{abs}\propto f$. On the other hand, the transverse components of the Sezawa-mode penetrates deeper into the material compared to the Rayleigh-mode~\cite{Cheeke2002}, leading to higher magnetoelastic driving field generated by the SAW in the YIG, which is buried beneath the ZnO layer in our samples. Thus, the SW excitation renders more efficient compared to the Rayleigh-type mode.\\
To model the obtained dataset, we follow the same approach discussed above for the Rayleigh-type mode. The FMR-excitation is computed as the $k=0$-mode with an out-of-plane excitation field, yielding an angle-independent excitation efficiency. The strain parameters of the SAW and the line-width are extracted using the same approach described prior, provided in Table\;\ref{tab:SAWFitparams}. The result is shown in Fig.\;\ref{fig:MECSezawaMode} b). We conclude that the effective field approach can sufficiently and accurately model the experimentally obtained data, for both for the Rayleigh-type mode and the Sezawa-type mode.

\section{Summary and Conclusions}

We presented a comprehensive study of the interaction of SAWs and SWs in a heterostructure of GGG/YIG/ZnO. The angle-dependence of the phonon-magnon-coupling was extraced by SAW transmission measurements. The measurements could be quantitatively replicated with our phenomenological model calculations based on magnetoacoustic interactions. By fitting our data to this model, we can extract the magnetic properties of the ferromagnetic material with high precision, as the interaction symmetry is strongly dependent on these parameters. Further, our data indicates that SWs in the ultra-low magnetic damping material YIG can be efficiently excited by magnetoacoustic interactions, as the SAW absorption is increased by a factor of 2.8\;dB at a frequency of 2.45\;GHz for the Rayleigh mode and up to 15\;dB at a frequency of 3.53\;GHz for the Sezawa mode. We note, that the total absorbed SAW power is highly dependent on the thickness composition of the material stack, as the characteristic decay of the SAW amplitude into the bulk varies strongly. Thus, by tuning and engineering the layer stack, the efficiency of the magnon pumping by the SAW can be further enhanced. The emergence of highly non-reciprocal coupling is characteristic to the inherent chirality in the system, by means of the intrinsic helicity of the SAWs and SWs.

\begin{acknowledgments}
This work was supported by the European Research Council (ERC) under the European Union’s Horizon Europe research and innovation programme (Consolidator Grant ``MAWiCS", Grant Agreement No. 101044526). The work of F. Ryburn was supported by a UK Engineering and Physical Sciences Research Council (EPSRC) Industrial Cooperative Award in Science \& Technology under Grant No. 2286081.
\end{acknowledgments}


\appendix

\section{Modeling of the magnetoelastic coupling}
\label{App:MEC}

We model the magnetic field and angle-dependent coupling of the SAW with SWs in the YIG by following the effectice field approach introduced by Dreher \andGang~\cite{Dreher2012} and Küß \andGang~\cite{kussNonreciprocalDzyaloshinskiiMoriya2020}. First, the equilibrium magnetization direction is computed by minimizing the free energy density $G$. For this, we consider the Zeeman-energy, the in-plane uniaxial anisotropy term and the out-of-plane anisotropy:
\begin{equation}
    G=-\mu_0\textbf{H}\cdot\textbf{M}+K_\text{u}(m_\text{x}m_\text{y}+m_\text{x}m_\text{z}+m_\text{y}m_\text{z})+K_\text{S}(\Tilde{n}\cdot\textbf{M})^2
\end{equation}
Here, $K_\text{u}$ denotes the uniaxial magnetic anisotropy constant, $\Tilde{n}$ is the demagnetizing tensor and $K_\text{S}$ the out-of-plane anisotropy constant. The dynamical driving field $\mu_0\textbf{h}_\text{MEC}$ generated by the SAW is derived from the magnetoelastic energy density $G^\text{d}$:
\begin{align}
    G^{\text{d}} =\;& b_{1}(\varepsilon_{\text{xx}}M_{\text{x}}^{2} + \varepsilon_{\text{yy}}M_{\text{y}}^{2} + \varepsilon_{\text{zz}}M_{\text{z}}^{2})  \notag\\
    & + 2b_{2} ( \varepsilon_{\text{xy}}M_{\text{x}}M_{\text{y}} + \varepsilon_{\text{xz}}M_{\text{x}}M_{\text{z}} + \varepsilon_{\text{yz}}M_{\text{y}}M_{\text{z}}).
\end{align}
The driving field $\mu_0\textbf{h}_\text{MEC}=(\mu_{0} h_{\text{1}}, \mu_{0} h_{\text{2}})^\text{T}$ is of the form:
\begin{align}
    \mu_{0} h_{\text{1}} =&  - 2\frac{b_{\text{2}}}{M_\text{S}} \big[\cos\phi_{\text{0}}\varepsilon_{\text{xz}} + \sin\phi_{\text{0}}\varepsilon_{\text{yz}}\big]\\
\mu_{0} h_{\text{2}} = & - 2\frac{b_{\text{2}}}{M_\text{S}}\cos(2\phi_{\text{0}})\varepsilon_{\text{xy}} +2\frac{b_{\text{1}}}{M_\text{S}} \sin\phi_{\text{0}}\cos\phi_{\text{0}} \varepsilon_{\text{xx}}.
\end{align}
Here, the "1" direction points parallel to the magnetization and the "2" direction is perpendicular in the OOP-direction. The response of the spin-wave system can then be computed by means of the Polder susceptibility $\chi$:
\begin{widetext}   
\begin{align}
    \textbf{M}&=\chi\textbf{h}_\text{MEC}\\
    \chi_{11}^{I} &= H \cos(\phi_0 - \phi_{\text{H}}) + \frac{2A}{\mu_0 M_{\text{s}}} k^2 + M_{\text{s}} G_0 - \frac{2K_\text{S}}{M_\text{S}} + \frac{2K_\text{u}}{M_\text{S}} \cos^2 (\phi_0 - \phi_{H}) - i \frac{\alpha_\text{eff} \omega}{\mu_0 \gamma}, \notag\\
    \chi_{12}^{I} &= -\chi_{21}^{I} = i \frac{\omega}{\mu_0 \gamma} , \notag\\
    \chi_{22}^{I} &= H \cos(\phi_0 - \phi_{\text{H}}) + \frac{2A}{\mu_0 M_{\text{s}}} k^2 + M_{\text{s}} (1 - G_0) \sin^2 (\phi_0) + \frac{2K_\text{u}}{M_\text{S}} \cos(2(\phi_0 - \phi_{H})) - i \frac{\alpha_\text{eff} \omega}{\mu_0 \gamma}\label{eq:PolderSus}.
\end{align}
\end{widetext}
Here $\omega=2\pi f$ is the angular frequency, $\alpha^\text{eff}=\frac{\gamma}{2\omega}\mu_0\Delta H+\alpha^\text{FMR}$ is the effective Gilbert damping, $A$ is the exchange constant and $M_\text{S}$ the saturation magnetization. $G_0$ describes the dipolar term by $G_0=(1-\exp(-\|kt\|)/(kd)$. The power absorption due to the excitation of the magnetic system via a dynamical rf-field is given by $P_{\text{SW}}=-\frac{\mu_0\omega}{2}\text{Im}\left(\int_V \textbf{h}^*\chi \textbf{h}\text{d}V\right)$~\cite{kobayashiFerromagnetoelasticResonanceThin1973}, where the integration is performed over the magnetic volume, under assumption of a uniform strain- and spin-wave-mode profile over the thickness of the magnetic layer. Due to the decay of the SAW the strength of the generated magnetoelastic driving field is reduced with increasing propagation. Consequently, the integral becomes a differential equation, which is solved by an exponential Ansatz~\cite{kussNonreciprocalDzyaloshinskiiMoriya2020}. Finally, the relative transmitted SAW power can be computed as:
\begin{align}
    \log\left(\frac{P_\text{SAW}}{P_0}\right)=\exp\left[-\frac{\omega\mu_0V}{2}\text{Im}(\textbf{h}_0^*\chi \textbf{h}_0)\right],
\end{align}
where $\textbf{h}_0$ refers to the driving field, generated by the SAW at the start of propagation.\\
Further, from the Polder susceptibility, we introduce the definitions $H_{11}=\text{Re}(\chi_{11})$ and $H_{22}=\text{Re}(\chi_{22})$, to obtain the spin-wave dispersion as:
\begin{equation}
    f=\frac{\gamma}{2\pi}\mu_0\sqrt{H_{11}H_{22}}.
\end{equation}
\begin{figure*}[t]
\includegraphics{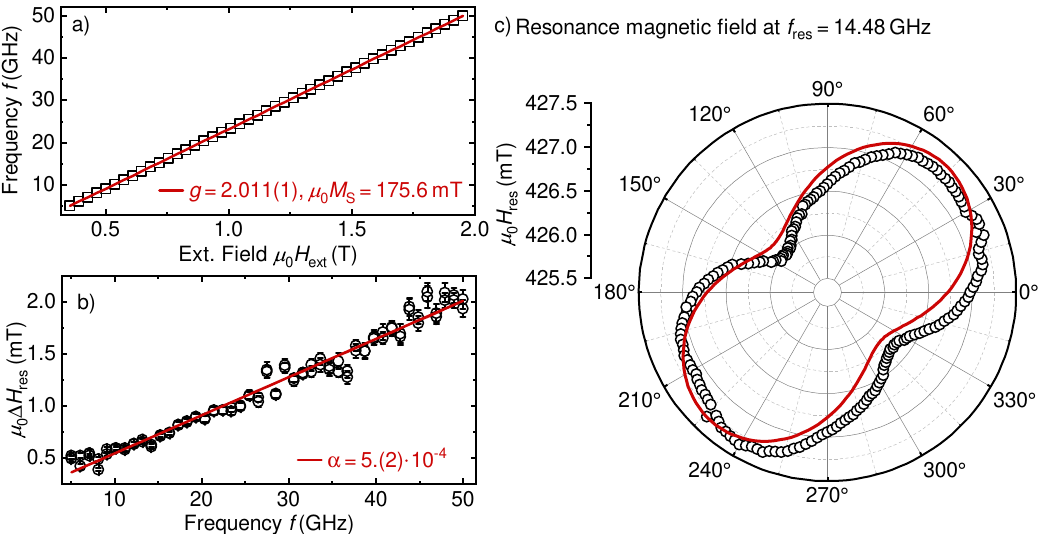}
\caption{\label{fig:FMR}Panel a) and b) show the example Kittel fit and Line-width result in the OOP-geometry, while c) shows the resonance magnetic field at a fixed frequency of $f=14.48\;$GHz. The orientation of the external field is defined via the angle $\phi_H$ relative to the wave vector of the SAW such that $\phi_H=0\|k_\text{SAW}$.}
\end{figure*}
\section{Magnetic parameters determinded by ferromagnetic resonance spectroscopy}
\label{App:FMR}

\begin{table}[b]
\caption{\label{tab:FMRParams}%
Magnetic parameters obtain from FMR spectroscopy measurement. The crystalline magnetic anisotropy constants $K_1$ and $K_2$ as well as the saturation magnetization $\mu_0M_\text{S}$ are taken from Lee \andGang~\cite{leeFerromagneticResonanceYIG2016}.}
\begin{ruledtabular}
\begin{tabular}{ll}
\textrm{Quantity}&
\textrm{Result}\\
\colrule
$g$ & 2.011 \\
$\mu_0M_\text{eff}$ (mT) & 183 \\
$\mu_0M_\text{S}$ (mT) & 176\textsuperscript{~\cite{leeFerromagneticResonanceYIG2016}} \\
$\alpha^\text{FMR}$ & $5.2\cdot10^{-4}$ \\
$\mu_0\Delta H_\text{inhomo}\;$(mT) & 0.17 \\
$K_\text{u}$ (J/m\textsuperscript{3}) & -51.0 \\
$\phi_\text{u}$ (°) & -43.3 \\
$K_\text{1}$ (J/m\textsuperscript{3}) & -542.1\textsuperscript{~\cite{leeFerromagneticResonanceYIG2016}} \\
$K_\text{2}$ (J/m\textsuperscript{3}) & -137.2\textsuperscript{~\cite{leeFerromagneticResonanceYIG2016}} \\
\end{tabular}
\end{ruledtabular}
\end{table}

We utilize well-established VNA-based broad-band ferromagnetic resonance spectroscopy to determine the magnetic parameters on the YIG upon ZnO-deposition. For this purpose, a blank, cut piece of the sample is used, to ensure the comparability. We perform FMR-measurements in the OOP-geometry to extract the saturation magnetization and the gyromagnetic ratio. The use of angle-resolved FMR spectroscopy is employed to determine the magnetic anisotropy. The result are shown in Fig.\;\ref{fig:FMR}.\\
We fit the OOP-FMR Data with the appropriate Kittel-Equation, given by:
\begin{equation}
    f_\text{res}(\mu_0H_\text{ext})=\frac{g\mu_\text{B}}{2\pi\hbar}(\mu_0H_\text{ext}-\mu_0M_\text{eff}).
\end{equation}
From the frequency-dependent line width, we determine the Gilbert-damping and the inhomogeneous line width, by fitting the expression:
\begin{equation}
    \mu_0\Delta H_{\text{res}}=\frac{2h\alpha}{g\mu_\text{B}}f+\mu_0\Delta H_{\text{inhomo}}
    \label{eq:LineWidth}
\end{equation}

We find a the Gilbert damping parameter of $\alpha=5.(2)\times10^{-4}$. This indicates that the low-magnetic damping property of the YIG is preserved during the ZnO-deposition. \\
In Fig.\;\ref{fig:FMR} c) we show the resonance magnetic field as a function of the orientation of the external magnetic field, examplary at a frequency of 14.48\;GHz. As can be seen, the six-fold anisotropy, expected for a cubic crystal structure oriented in [111]-direction, is overlapped by a dominat uniaxial anisotropy. To determine the magnetic anisotropy parameters, we calculate the angle-dependent ferromagnetic resonance frequency as follows. We apply an adapted approach inspired by Lee \andGang~\cite{leeFerromagneticResonanceYIG2016} and start by from the free energy density $G$~\cite{RezendeBook}:
\begin{figure*}[t]
\includegraphics{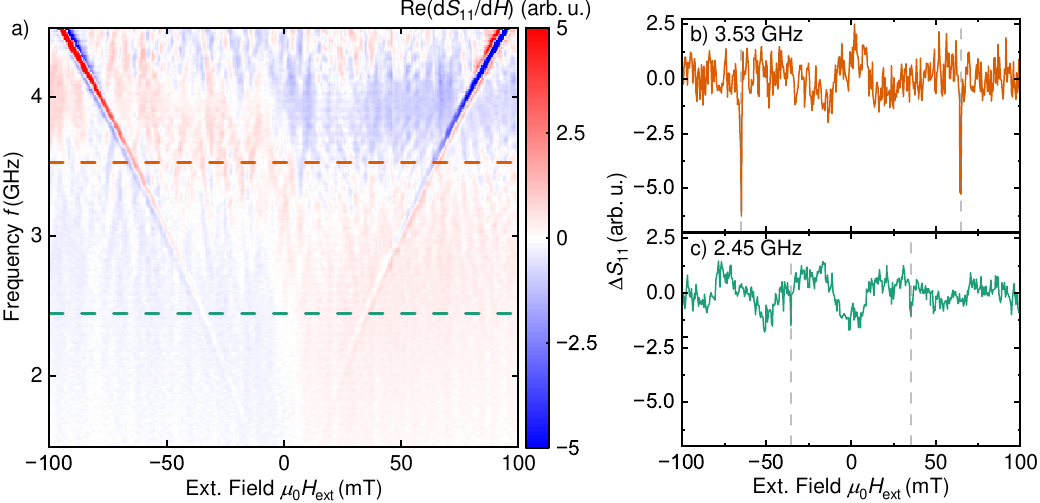}
\caption{\label{fig:S11Reflec}$S_{11}$-Reflection Measurement as a function of external magnetic field $\mu_0H_{\text{ext}}$ with one of the IDTs connected to the VNA. a) Shows the reflected $S_{11}$-signal, derived with respect to the external magnetic field, revealing the FMR signature. The green line at 2.45\;GHz and the orange line at 3.53\;GHz highlight the frequency at which the Rayleigh- and Sezawa-type SAWs are excited, respectively. In Panel b) the background-corrected $S_{11}$-signal is shown at 2.45\;GHz as a function of the external magnetic field, while Panel c) shows the background-corrected signal at 3.53\;GHz. At their certain FMR-fields the additional absorption due to the excitation of the FMR become apparent.}
\end{figure*}

\begin{align}
    G=&-\mu_0\textbf{H}\cdot\textbf{M}+\frac{1}{2}\mu_0(\Tilde{n}\cdot\textbf{M})^2+\\
    &K_1(m_\text{x}^2m_\text{y}^2+m_\text{y}^2m_\text{z}^2+m_\text{z}^2m_\text{x}^2)+K_2m_\text{x}^2m_\text{y}^2m_\text{z}^2+\notag \\
    &K_\text{u}(m_\text{x}m_\text{y}+m_\text{x}m_\text{z}+m_\text{y}m_\text{z})+K_\text{S}(\Tilde{n}\cdot\textbf{M})^2.\notag
\end{align}

We use the normalized magnetization $\textbf{m}=\textbf{M}/M_\text{S}$. $K_\text{1}$ and $K_\text{1}$ denote the first and second order cubic magnetic anisotropy constants, $K_\text{u}$ the uniaxial anisotropy constant and $K_\text{S}$ the out-of-plane anisotropy constant. The resonance frequency is derived using the famous relation by Suhl~\cite{Suhl1955} and Smith and Beljers~\cite{smit1955} $f=\frac{\gamma}{2\pi M_\text{S}\sin\theta}(G_{\theta\theta}G_{\phi\phi}-G_{\theta\phi}^2)^{1/2}$. As we measure in the high-field limit the magnetic field dragging effect is neglected. We compute the angle-dependent FMR-frequency as:

\begin{widetext}
\begin{align}
    f(H_0,\phi_\text{H})=\frac{g\mu_\text{B}}{h}\cdot&\sqrt{\left(\mu_0H_0+\mu_0M_\text{eff}+\frac{2K_\text{u}}{M_\text{S}}\cos^2(\phi_0-\phi_\text{u})-\frac{K_1}{M_\text{S}}+\frac{K_2}{18M_\text{S}}(2-\cos(6\phi_\text{H})\right)\cdot}\notag\\
    &\overline{\left(\mu_0H_0+\frac{2K_\text{u}}{M_\text{S}}\cos(2(\phi_0-\phi_\text{u}))-\frac{K_2}{3M_\text{S}}\cos(6\phi_\text{H})\right)}.\label{Eq:FMR}
\end{align}
\end{widetext}
As is common, in Eq.\;\eqref{Eq:FMR} we replaced the arising term $\mu_0M_\text{S}-\frac{2K_\text{S}}{M_\text{S}}$ with the effective magnetization $\mu_0M_\text{eff}$. The obtained magnetic parameters are summarized in Table\;\ref{tab:FMRParams}.\\
Investigating the origin of the dominating uniaxial is out of scope of this study. However we suspect it to originate from the anisotropic thermal expansion of the YIG, which is expected in Wurzite-type crystals~\cite{IwanagaWurzitCrystals2000}. During the ZnO sputtering process, the substrated is heated, thus the different thermal expansion coefficient of the YIG and ZnO induce a permenant strain on the YIG when cooling down, resulting in an additional uniaxial anisotropy. Such observations were reported before~\cite{kussNonreciprocalDzyaloshinskiiMoriya2020}.\\
As is apparent from the angle-resolved FMR-measurement shown in Fig.\;\ref{fig:FMR}, the uniaxial anisotropy vastly dominates over the crystalline anisotropy, while still being in the mili-Tesla range. Consequently, for modeling the magnetoelastic interaction, we consider only the terms associated with the uniaxial anisotropy.

\section{Direct FMR-excitation by microstructured pads}
\label{App:S11}

As stated in the main text, we observe in Fig.\;\ref{fig:MECSezawaMode} a) a direct excitation of the ferromagnetic resonance. Since the excitation of the FMR via the SAW can be excluded due to the violation of the momentum conservation, we suspect the additional absorption line to originate as follows. In commonly employed devices, the magnetic film is deposited between the IDTs on a piezoelectric layer, while the microstructured IDTs are located only on the piezoelectric~\cite{Kuess2024}. In our device the IDTs are placed on top of the piezoelectric layer which covers the YIG. The alternating current in the microstructured patterns generates an approximately uniform magnetic field, which directly interacts with the magnetic system in the YIG and excites its FMR. To test this hypothesis we measured the reflected signal $S_{11}$ as a function of field and frequency, as is shown in Fig.\;\ref{fig:S11Reflec}.\\
We apply the well-established derivative divide method~\cite{maierflaigNoteDerivativeDivide2018} to extract the magnetic field dependent signal and remove the frequency-dependent background. As illustrated in Fig.\;\ref{fig:S11Reflec} a), the excitation of the ferromagnetic resonance is visible. Notably, it occurs continuously in frequency. As such, it is not constrained by IDT periodicity, which would otherwise permit the the excitation of only discrete wave-vectors, and consequently, frequencies. Nevertheless, the SAW excitation remains visible, as illustrated by the line cuts shown in Fig.\;\ref{fig:S11Reflec} b) at the frequency corresponding to the excitation of the Rayleigh-type mode at $2.45\;$GHz. A similar observation holds for the frequency of $3.53\;$GHz, indicating the excitation of the Sezawa-type mode in Fig.\;\ref{fig:S11Reflec}.\\
We argue that due to the direct excitation of the FMR, less microwave power applied to the IDT is available for the SAW excitation. This results in an additional field-dependent, but angle-independent, absorption line in the SAW-transmission measurement, as observed in Fig.\;\ref{fig:MECSezawaMode}. Further, the FMR-line is clearly visible only for the Sezawa-mode at 3.53\;GHz, and not at 2.45\;GHz, because generally the microwave power absorption scales with the frequency, following the relation $P_\text{abs}\propto f$~\cite{RezendeBook}.

\nocite{*}

\bibliography{Bibliography}

\end{document}